\documentclass[aps,pra,reprint,superscriptaddress,longbibliography]{revtex4-1}

\usepackage{graphicx}% Include figure files
\usepackage{dcolumn}% Align table columns on decimal point
\usepackage{bm}
\usepackage{amsfonts}
\usepackage{amssymb}
\usepackage{amsmath}
\usepackage{graphicx}
\usepackage{color}
\usepackage{dcolumn} % Align table columns on decimal point
\usepackage{bm} % bold math
\usepackage{amsthm}
\usepackage{latexsym}
\usepackage{float}
\usepackage{tabularx}

\begin{document}

\title{Characterizing pump line phase offset of a single-soliton Kerr comb by dual comb interferometry}

\author{Ziyun Kong,$^{1, \dag, *}$ Chengying Bao,$^{1, 4, \dag}$  Oscar E. Sandoval,$^{1}$ Bohao Liu,$^{1}$ Cong Wang,$^{1}$ Jose A. Jaramillo-Villegas,$^{1, 2}$ Minghao Qi,$^{1, 3}$ Andrew M. Weiner$^{1, 3}$\\
\vspace{1mm}
$^1$ School of Electrical and Computer Engineering, Purdue University, 465 Northwestern Avenue, West Lafayette, Indiana 47907-2035, USA\\
$^2$ Facultad de Ingenierías, Universidad Tecnológica de Pereira, Carrera 27 \#10-02, Pereira, Risaralda 660003, Colombia\\
$^3$ Birck Nanotechnology Center, Purdue University, 1205 West State Street, West Lafayette, Indiana 47907, USA \\
$^4$ T. J. Watson Laboratory of Applied Physics, California Institute of Technology, Pasadena, California 91125, USA.\\
$^\dag$These authors contributed equally to this work.\\
$^*$Corresponding author: kongz@purdue.edu
}

\begin{abstract}
We experimentally demonstrate phase retrieval of a single-soliton Kerr comb using electric field cross-correlation implemented via dual-comb interferometry.   The phase profile of the Kerr comb is acquired through the heterodyne beat between the Kerr comb and a reference electro-optical comb with a pre-characterized phase profile. The soliton Kerr comb has a nearly flat phase profile, and the pump line is observed to show a phase offset which depends on the pumping parameters. The experimental results are in agreement with numerical simulations. Our all-linear approach enables rapid measurements (3.2 $\mu$s) with low input power (20 $\mu$W).
\end{abstract}

\maketitle

Kerr combs, frequency combs generated by externally pumping high-Q microresonators with a continuous-wave (CW) laser, have seen considerable attention in the last decade \cite{Kippenberg_Science2018}.  These compact, high repetition rate, and potentially CMOS compatible comb sources are of interest for a wide range of applications, including optical communications \cite{Kippenberg_Nature2017Commun}, optical arbitrary waveform generation \cite{Weiner_NP2011}, and spectroscopy \cite{Vahala_Science2016}. Soliton generation, which occurs in anomalous dispersion microresonators, is an importance mechanism for generation of low noise, broadband Kerr combs with smooth spectra \cite{Kippenberg_Science2018,Kippenberg_NP2014,Vahala_Optica2015,Gaeta_Ol2016Thermal,Weiner_OE2016}.  In contrast to frequency combs based on mode-locked lasers, the CW pump is coherently coupled to the Kerr comb.  An approximate analytic solution for the electric field amplitude of single soliton Kerr comb in the framework of the standard Lugiato-Lefevre equation may be written \cite{Kippenberg_NP2014,Wabnitz_OL1993}

\begin{equation}
    a(t)=a_0+Ae^{i\phi_0}\textrm{sech}(t/t_p)
\end{equation}
where $a_0$ is a CW background field, and $A$, $\phi_0$ and $t_p$ are the amplitude, phase shift, and pulse width parameter of the soliton. (This field is assumed to repeat periodically with the cavity round trip time.) The complex spectrum consists of a strong line at the pump frequency, superimposed on and phase shifted with respect to the smooth spectral envelope of the soliton.  The phase shift of the pump line plays a key role in mediating the parametric gain affecting the comb and has been shown to be necessary based on arguments from self-organization theory \cite{Gaeta_PRA2016}. The existence of such a phase shift has been demonstrated experimentally via a method based on pulse shaping and intensity autocorrelation, both for soliton Kerr combs \cite{Weiner_OE2016} and dark pulse Kerr combs in normal dispersion microresonators \cite{Weiner_NP2015}.

However, these measurements which rely on optical nonlinearity are difficult and time consuming, especially for soliton Kerr combs which have low power.  In this Letter we demonstrate an all-linear method for phase retrieval of soliton Kerr combs using electric field cross correlation (EFXC) \cite{Weiner_OL2009}, a technique which shares the same technical principle as dual comb spectroscopy \cite{Newbury_Optica2016}. This method delivers rapid waveform measurement and high sensitivity for measurement at low input power. High quality retrieval of the spectral phase is achieved with 20 $\mu$W average power from a 227.5 GHz repetition rate Kerr comb, corresponding to $<$0.1 fJ per pulse. A similar technique was recently used to capture rapid scenes in microresonators such as soliton breathing dynamics, but the phase information of the stable solitons was not studied \cite{Vahala_NC2018,Kippenberg_NP2018}. Our experiments confirm that soliton Kerr combs feature a pump line with substantial phase offset with respect to the soliton spectrum which otherwise has nearly flat spectral phase.  Moreover, the improved measurement capability now allows us to reveal a significant dependence of the phase offset on pump power and detuning.  Numerical simulations based on the Lugiato-Lefever equation (LLE) \cite{Coen_OL2013} show trends similar to those obtained in experiment.

\begin{figure}[t]
\centering
\includegraphics[width=\linewidth]{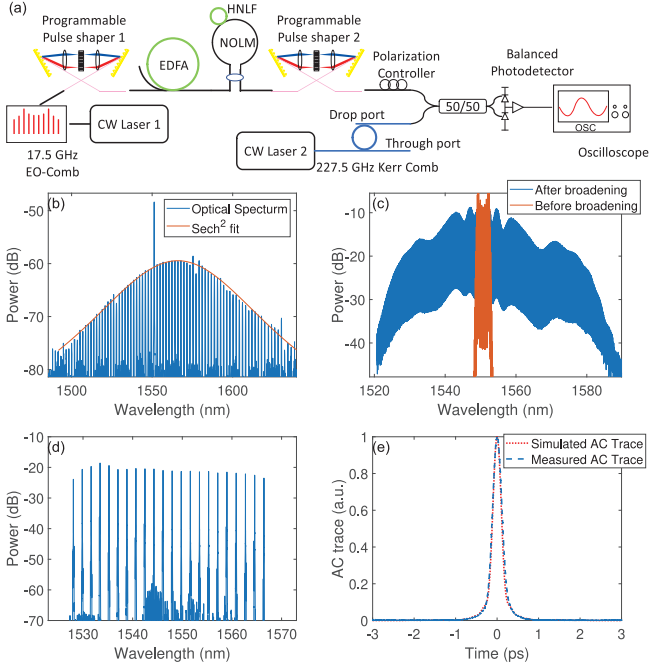}
\caption{(a) Experimental setup and dual comb sources. The phase of the soliton Kerr comb is measured by an EO comb with an offset repetition rate. Pulse shaper 2 is used to shape the EO comb into a transform-limited pulse with approximately flat phase profile. CW: continuous wave; EO comb: electro-optic comb; EDFA: Erbium-doped fiber amplifier; NOLM: nonlinear optical loop mirror; HNLF: Highly-nonlinear fiber. (b) Spectrum of the soliton comb; (c) Spectrum of EO comb before and after broadening; (d) Spectrum of broadened EO comb after filtering to suppress the unused comb lines and flattening; (e) Autocorrelation trace of compressed EO comb with comparison to simulated trace.}
\label{fig:setup}
\end{figure}

\begin{figure}[t]
\centering
\includegraphics[width=\linewidth]{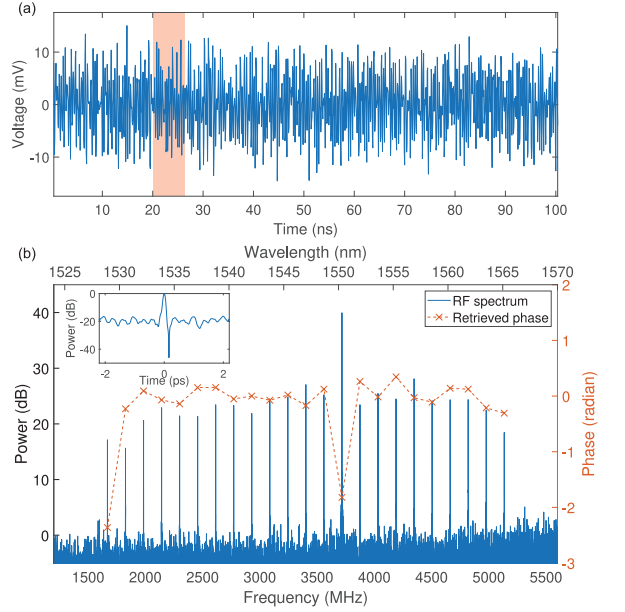}
\caption{(a) Portion of time domain interferogram recorded by an oscilloscope. The shaded box represents one period of the interferogram. (b) Blue line is the power spectrum of the interferogram. Orange line is the retrieved phases for different optical comb modes, showing a phase offset for the pump line. The inset is the reconstructed intracavity waveform using the measured comb phase, showing a dip at the tail of the pulse.}
\label{fig:phase}
\end{figure}

\begin{figure}[t]
\centering
\includegraphics[width=\linewidth]{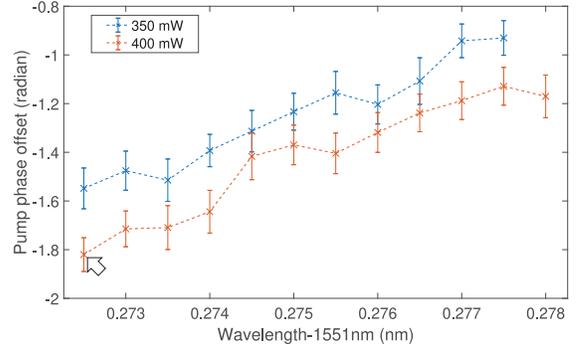}
\caption{Retrieved pump line phase offset with increasing pump wavelength (detuning). The point indicated by the arrow corresponds to the data in Fig 2(c).}
\label{fig:result}
\end{figure}

\begin{figure}[t]
\centering
\includegraphics[width=\linewidth]{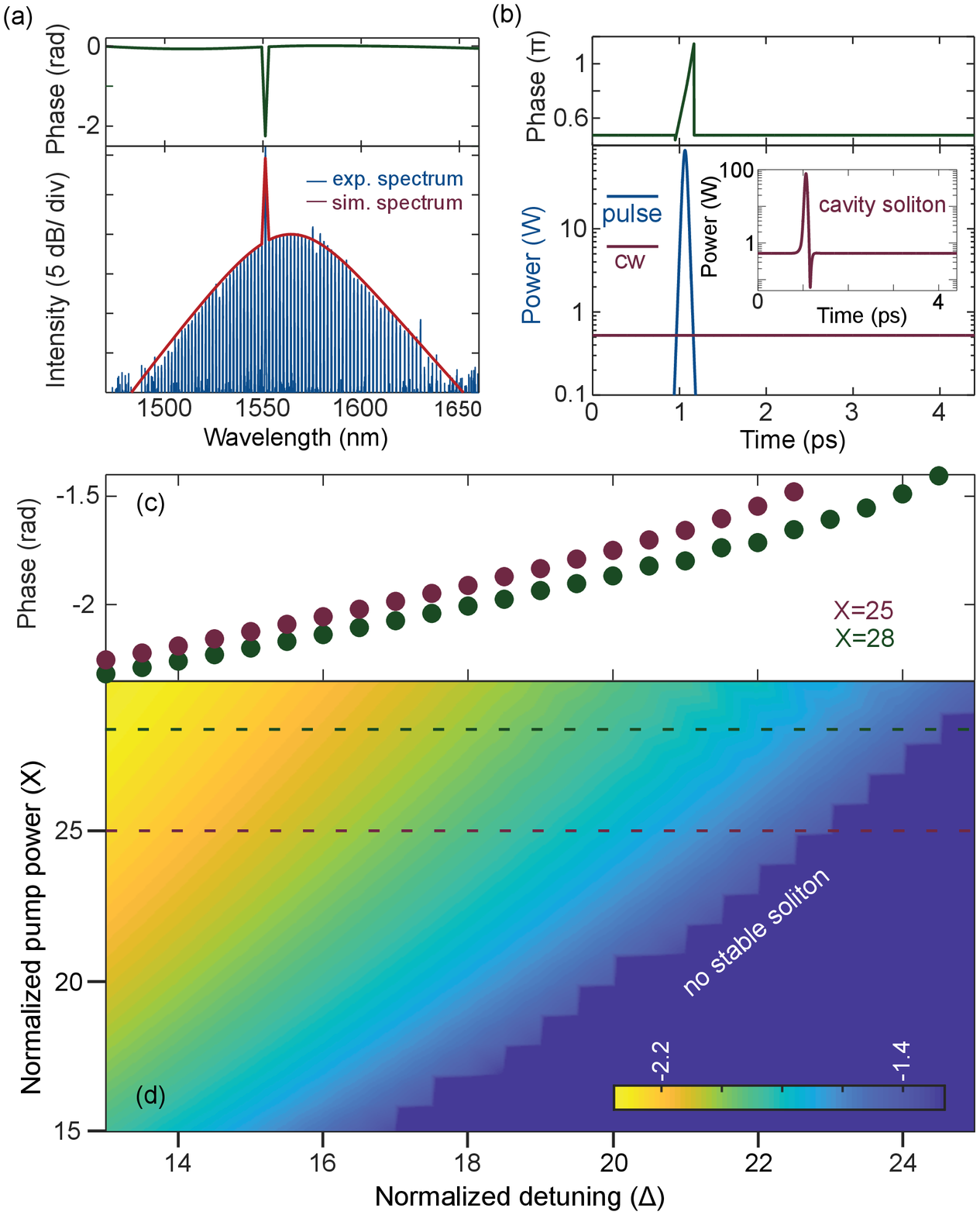}
\caption{(a) An example of the simulated spectrum (red) and experimental spectrum (blue) of the soliton Kerr comb with $X$=25, $\Delta$=13. The simulated intracavity soliton also has a nearly flat phase profile with the pump line exhibiting a negative phase offset. (b) The pulse-part and background-part of the cavity soliton have different phase and their destructive interference can lead to a dip of the simulated waveform (inset). The top panel shows the phase difference between the pulse-part and the cw-background-part. (c) Pump phase offset becomes less negative with increasing detuning for a fixed pump power (X=28 and X=25 represented by the horizontal dashed lines in (d)). (d) Change of the pump phase offset versus pump power and detuning. There is no stable soliton in the dark blue region.}
\label{fig:sim}
\end{figure}

The experimental setup is illustrated in Fig. 1(a). The system includes two combs (signal comb and reference comb) which differ slightly in repetition rates. Due to the repetition rate difference between the two combs, the reference pulse will sweep across the signal pulse automatically to generate the EFXC signal without any mechanical scan. Equivalently, in the frequency domain, two frequency combs with a slightly different repetition rate heterodyne beat to generate a radio frequency (RF) comb. The phase of the RF comb $\phi_{\textrm{RF}}^i= \phi_{\textrm{sig}}^i-\phi_{\textrm{ref}}^i$ (the superscript $i$ refers to the $i$th comb line) can be measured by recording the EFXC interferogram. Therefore, the phase of the signal comb ($\phi_{\textrm{sig}}^i$) can be obtained once the reference comb phase ($\phi_{\textrm{ref}}^i$) is known. In our experiments, the signal comb is generated from a silicon nitride microresonator (radius 100 $\mu$m and loaded-Q 2.4 million) with a repetition rate of $\sim 227.5$ GHz \cite{Weiner_OE2016,Weiner_OL2017RepRate}. To avoid the complication of strong, directly transmitted pump field superimposed on top of the comb at the through port, we use a drop port which provides a direct sample of the intracavity comb field \cite{Weiner_OE2016}. However, because the drop port is designed to have relatively low coupling in order to minimize reduction of the microresonator Q-factor, the output power is reduced. The spectrum of the comb sampled at the drop port, Fig. 1(b), comprises a $\textrm{sech}^{2}$-like spectral envelope together with an $\sim 10$ dB stronger pump line corresponding to the weak background accompanying the soliton. Since the repetition rate of the signal comb is high, we use an electro-optical (EO) comb, which also has a relatively high repetition rate, as the reference comb \cite{Weiner_STJQE2013}. The spectrum of the as generated EO comb is shown as the orange line in Fig. 1(c).  A first pulse shaper is used to compress the chirped EO comb into a pulse train. The pulses are then amplified and spectrally  broadened within a nonlinear optical loop mirror (NOLM) \cite{Radic_JLT2014} constructed using highly nonlinear fiber, resulting in spectra that  span over 50 nm around 1550 nm (blue line in Fig. 1(c)). The NOLM also cleans the pedestal of the associated pulses in the time domain. Since the EO comb is driven by a 17.5 GHz microwave synthesizer (i.e., its repetition rate is 17.5 GHz), only one out of every 13 lines of the EO comb will beat with the soliton Kerr comb to generate low frequency signals that can be detected by the photodetector. The unused EO comb lines are filtered out by a second pulse shaper to avoid associated excess noise contributions (see Fig. 1(d) for the filtered EO comb). The second pulse shaper is also used in line-by-line pulse shaping mode \cite{Weiner_NP2011,Weiner_NP2015,Weiner_OL2018} to equalize the intensities of the remaining comb lines and compress them into a transform-limited pulse train at 227.7 GHz, with $\sim$ 160 MHz repetition rate offset compared to the Kerr comb.  Figure 1(e) depicts the autocorrelation of the compressed pulses. The close agreement between the measurement and the simulated autocorrelation using the known power spectrum with the assumption of flat spectral phase provides evidence that the compressed pulses are close to transform-limited; hence we can approximate $\phi_{\textrm{ref}}\approx 0$.

The input powers (pulse energies) for our dual comb interferometry measurements are typically set at 20 $\mu$W ($\sim0.1$ fJ) and 1 mW ($\sim5$ fJ) for the soliton Kerr comb and the EO comb, respectively. The interferogram is detected by a balanced photodetector (Discovery Semiconductor DSC720: balanced InGaAs photodiodes to 20 GHz) and recorded by an oscilloscope (Tektronix DSA72004B: digital serial analyzer with 20 GHz analog bandwidth and running at 12.5 GHz sampling rate, see Fig. 2(a)).   Due to the dispersion of the fiber connection between the microresonator drop-port and the 50/50 fiber coupler ($\sim7.5$ m of Corning\textsuperscript{\small{\textregistered}} SMF-28e fiber), the RF beat signal is broadened in time, and contributions from adjacent Kerr comb solitons blend together.  The dispersion of the fiber length is characterized and yields a value of 0.135 ps/nm, consistent with the length of fiber.  The phase resulting from the fiber dispersion is subtracted from the Kerr comb spectral phase obtained from our measurements. The Fourier transform of a 3.2 $\mu$s duration interferogram (corresponds to $\sim 512$ periods of the EXFC signal) yields an RF comb, with power spectrum shown in Fig. 2(b). 23 lines are visible, corresponding to an optical bandwidth of $\sim5$ THz, limited by the passband of pulse shaper 2, used for compression of the broadened reference comb. The line corresponding to the pump has an SNR exceeding 40 dB; the other lines have SNRs ranging between 17 dB and 30 dB.  The phase of the single soliton Kerr comb is shown in Fig. 2(b). The phase of the pump line is shifted by approximately $-$1.8 rad  with respect to the rest of the lines, for which the phase is approximately flat (standard deviation $\sim 0.19$ radian) - as expected for a single intracavity soliton. In the time domain this corresponds to a soliton with a positive phase shift with respect to the background, consistent with the approximate analytical solution discussed in \cite{Wabnitz_OL1993,Kippenberg_NP2014}.  Note that the signs of all phases are defined assuming an $exp(-i\omega t)$ carrier, as in the standard form of the Lugiato-Lefevre equation (LLE).  A phase shift is also observed for the 1528 nm comb line, the shortest wavelength within our measurement range.  This may be related to mode-interaction induced Cherenkov radiation, also termed dispersive wave emission (see variations around 1528 nm in the soliton spectrum shown in Fig. 1(b)) \cite{Matsko_OL2016,Weiner_Optica2017}. This phase offset suggests that the Cherenkov radiation also has a phase difference compared to the soliton.  Using the measured comb phase, we are able to reconstruct the intracavity waveform (comb lines outside the passband of the pulse shaper 2 are assumed to have zero phase with powers that follow the sech$^2$ fit), see inset of Fig. 2(b). There is a dip at the tail of the pulse, due to the pump phase shift (a similar effect is seen in the simulations of Fig. 4). These results demonstrate that all-linear dual comb interferometry enables characterization of the soliton Kerr comb at low power with high acquisition speed and clearly reveal the phase shift of the soliton with respect to the pump.

The rapid measurement offered by dual comb interferometry allows us to study the dependence of the pump phase offset on pumping conditions. The pump phase offset at a given pumping condition is measured through averaging the retrieved phase of 100 independently captured interferograms. The error bars, taken from the standard deviation among retrieved phase profiles with linear and constant terms removed, are $\sim0.08$ radian.  Measurement results are shown in Fig. 3 for on-chip pump powers of 350 mW and 400 mW (estimated based on 3 dB fiber-to-chip coupling loss) for the full range of pump wavelengths over which stable single solitons were obtained.  For fixed 400 mW pump power, the phase offset is observed to change from $-$1.8 rad to $-$1.2 rad with increasing pump wavelength (corresponding to increasing pump detuning, $\delta_0$, defined below). Moreover, the magnitude of the pump phase offset is larger with increased pump power.   From the power spectra and pump phase offsets, we can estimate the behavior in the time domain.  The ratio of the soliton peak power to the background is of order 100 at small detuning and increases by roughly a factor of two as the detuning is increased. The phase of the soliton with respect to the background is positive; the magnitude of the soliton phase offset with respect to the background shows trends similar to Fig. 3 (i.e., magnitude of the soliton phase offset decreases with increasing detuning). These trends are consistent with the behavior predicted by the approximate analytical solution \cite{Kippenberg_NP2014,Wabnitz_OL1993}

The measurement results are in reasonable agreement with numerical simulation based on the generalized LLE with Raman effect included, which can be written as \cite{Coen_OL2013,Weiner_PRL2016}

\begin{equation}
\begin{aligned}
& \left( {{\tau }_{R}}\frac{\partial }{\partial t}+\frac{{{\alpha}}+\theta }{2}+i{{\delta }_{0}}+i\frac{{{\beta }_{2}}L}{2}\frac{{{\partial }^{2}}}{\partial {{\tau }^{2}}} \right)E-i(1-f_{R})\gamma L{{\left| E \right|}^{2}}E \\
& -if_{R}{{\gamma }}L\left( E\int_{-\infty }^{\tau }{{{h}_{R}}\left( \tau -\tau ' \right){{\left| E \right|}^{2}}d\tau '} \right)-\sqrt{\theta }{{E}_{in}}=0, \\
\end{aligned}
\label{EqLLE}
\end{equation}
where $E$ is the envelope of the intracavity field, $\tau_R$ is the round-trip time (4.4 ps), $L$ is the cavity length (628 $\mu$m), $\tau$ and $t$ are the fast and slow time respectively, $\alpha$ and $\theta$ are the intrinsic loss and the external coupling coefficient respectively, $\beta_2$ is the group velocity dispersion, $\gamma$ is the nonlinear coefficient, $|E_{in}|^2$ is the pump power, $\delta_0 = (\omega_o-\omega_p)\tau_R$ is the pump detuning, expressed as a round trip phase shift ($\omega_o$ is the resonance frequency and $\omega_p$ is the pump frequency), and $f_R$ is the Raman fraction. $h_R(\tau)$ is the Raman response function, which is calculated in the frequency domain \cite{Weiner_PRL2016}. The Raman effect is assumed to have a Lorentzian gain spectrum, whose peak is centered at $-$14.3 THz and bandwidth is 2.12 THz. For simplicity, we normalize the pump power and detuning as $X = 8|E_{in}|^2\gamma\theta L/(\alpha + \theta)^3, \Delta = 2\delta_0/(\alpha + \theta)$.

For simulations we choose $\alpha=0.0024$, $\theta=0.0011$, $\beta_2 = -61 \textrm{ps}^2/\textrm{km}$, $\gamma=0.9 \textrm{W}^{-1}\textrm{m}^{-1}$, and $f_R=0.13$, which we believe are representative of the experimental values. As an example, a stable soliton can be generated by setting the pump power and detuning to $X=25$ (240 mW) and $\Delta=13$ (frequency detuning 820 MHz. phase detuning, $\delta_0$=0.023). The simulated spectrum of a single soliton, Fig. 4(a), is in good agreement with the measured spectrum. The overall phase profile of the simulated comb is also nearly flat, and there is an offset for the pump line ($-$2.2 rad).  The offset is slightly larger than experiments, which may reflect uncertainties in the experimental parameters used for the simulation. The phase offset of the pump line affects the intracavity time domain waveform, resulting in a dip near the tail of the soliton (inset of Fig. 4(b), similar to Fig. 2(b) inset). By separating the intracavity soliton into the pulse-part and the cw-background-part (see eq. 1) and analyzing their phase difference (phase of the pulse-part minus that of the cw-background-part), it can be found that their destructive interference leads to this dip (see Fig. 4(b)). Note that the phase of the pulse-part is not evaluated for power less than 0.01 W and the constant phase in the top-panel is from the phase of the cw-background-part. The waveform is asymmetric (only one dip) as the Raman effect red-shifts the soliton center frequency, causing a linear phase across the pulse-part; thus, destructive interference only occurs at one side of the pulse.  If the Raman effect is taken out of the simulation, the waveform is symmetric with dips on both sides. We performed further simulations in which we varied the normalized pump power $X$ and detuning $\Delta$ to study how the phase offset of the pump changes with pump parameters. The results are shown in Fig. 4(d) as a function of both $X$ and detuning $\Delta$. In general, we find the phase offset become less negative with increasing pump detuning and decreasing pump power. Figure 4(c) shows two examples of the simulated phase offset versus detuning for fixed pump powers. Furthermore, comparison between the cases with and without Raman effects shows no significant difference in the pump phase shift. The simulation results are in reasonable agreement with the measurement results in Fig. 3. Qualitatively, we can understand these results if we view the comb line at the pump frequency as arising from the coherent addition of a CW background contribution and a contribution from the soliton spectrum evaluated at the pump frequency.  With an increase in detuning, the contribution from the background part decreases relative to that from the soliton.  Since the contribution from the soliton at the pump frequency has the same phase as its contributions to other comb lines, the overall phase offset of the pump line with respect to other comb lines tends to decrease as the detuning increases.

In conclusion, we have demonstrated phase retrieval of a low power soliton Kerr comb using dual comb interferometry.  Coupling from the drop port gives us a direct replica of the intracavity pump field and enables the study of the phase for the pump line.  With respect to the rest of the spectrum, the pump line is found to have a negative phase offset which becomes more negative with increasing pump power or decreasing pump detuning. These measurements are consistent with numerical simulations based on the generalized LLE.   Because this   method is based on linear optics and does not require high peak power, it is especially useful for characterizing signals combs which have very high repetition rates corresponding to low pulse energies.

\bigskip
\noindent\textbf{Funding} The work was supported in part by NSF under grant 1509578-ECCS, by AFOSR grant FA9550-15-1-0211, and by DARPA under grant W31P4Q-13-1-0018.

\vspace{3mm}
\noindent\textbf{Acknowledgment} CB gratefully acknowledge a postdoctoral fellowship from the Resnick Institute, Caltech.

\bibliography{sample}

\end{document}